\documentclass{article}

\usepackage[top=1.2in, bottom=1in, left=1in, right=1in]{geometry}
\usepackage{hyperref}
\usepackage{comment}
\usepackage{epsfig}
\usepackage{amssymb}
\usepackage{amsmath}
\usepackage{amsfonts}
\usepackage{color}

\newcommand{\eval}{\operatorname{Evaluate}}

\newcommand{\mySpOne}{\hspace*{0.1in}}
\newcommand{\mySpTwo}{\mySpOne\mySpOne}

\begin{document}
\title{Enabling Secure Database as a Service using Fully Homomorphic Encryption: Challenges and Opportunities}
\author{
Murali Mani, Kinnari Shah, Manikanta Gunda \\
University of Michigan, Flint \\
\{mmani, kishah, mgunda\}@umflint.edu
}
\date{\today}

\maketitle

\begin{abstract}
The database community, at least for the last decade, has been grappling with querying encrypted data, which would enable secure database as a service solutions. A recent breakthrough in the cryptographic community (in 2009) related to fully homomorphic encryption (FHE) showed that arbitrary computation on encrypted data is possible.  Successful adoption of FHE for query processing is, however, still a distant dream, and numerous challenges have to be addressed. One challenge is how to perform algebraic query processing of encrypted data, where we produce encrypted intermediate results and operations on encrypted data can be composed. In this paper, we describe our solution for algebraic query processing of encrypted data, and also outline several other challenges that need to be addressed, while also describing the lessons that can be learnt from a decade of work by the database community in querying encrypted data.
\end{abstract}

\section{Introduction}
There is significant interest among end users as well as enterprises in moving data and computation to the cloud. For instance, tools such as Google Docs, Microsoft Office 365, Turbo Tax Online help end users access the latest software using just a browser (typically) without worrying about installation, hardware capability,  operating system etc. Enterprises can get started without purchasing their expected future software and hardware needs~\cite{Huff:Li:10}. An example is the gaming company FarmVille (cited in~\cite{BOOK:ELLS:12}), which uses Amazon Web Services (AWS) and found it easy to grow from 1 million users in four days to 10 million users in two months to 75 million users in nine months. Examples of companies successfully using AWS are listed at {\url{https://aws.amazon.com/solutions/case-studies/}} including NASA/JPL for streaming live video of the landing of the Mars rover Curiosity.


Providing software as a service on the cloud is beneficial to software developers as well. 
The software provider can choose one environment on which the software will be deployed in the cloud and not worry about multiple platforms; also, upgrades to software by developers is easy~\cite{BOOK:ELLS:12}.

It would be remiss not to mention the security concerns in moving to the cloud. A recent example is the hacking of the account of Mat Honan, a technology writer~\cite{WIRED:Honan:12} on Aug 3, 2012, where hackers were able to break into his Google, Twitter and AppleID accounts, and remotely erase all of the data on his iPhone, iPad and MacBook. The above hack started by exploiting a contradiction in the security policies of Apple and Amazon to reset his appleID account, then the appleID account was used to get into Google and Twitter accounts, and iCloud's ``Find My" tool was used to remotely wipe his iPhone, iPad and MacBook. 

Preventing such hacks requires careful analysis of the vulnerabilities due to integration of services from multiple service providers, and is beyond the scope of this work. Instead, in this paper, we investigate how we can utilize the cloud for secure data management, where the cloud service provider stores the data and performs processing on the data in a secure manner. Currently, several companies provide document and file storage such as {\url{http://www.dropbox.com}}, which uses Amazon S3 {\url{http://aws.amazon.com/s3/}} for storage. Amazon recommends that the users encrypt their sensitive data before uploading it to S3. Amazon also provides simpleDB {\url{http://aws.amazon.com/simpledb/}} 
for storing data in a database as well as for processing of data. Amazon recommends that sensitive data stored in simpleDB also be encrypted by the client; however, if data is encrypted, then no processing can be performed on that data.{\footnote{CipherCloud {\url{http://www.ciphercloud.com/}} is more in line with our focus in this paper, and claims to provide format and function preserving AES compatible encryption schemes; however details are not available.}}


We can generalize the state of the art commercial solutions as follows. First, all communication and exchange of data (both within the cloud service provider and with the outside world) can be secured using encryption; however, within any component, processing will be performed on unencrypted data. Secondly, the client can encrypt the data and expose only the encrypted data to the service provider; however, little processing can be performed by the service provider on encrypted data. Unencrypted data within a cloud service provider is not secure, even when the service provider is ``trusted". For instance,~\cite{CCS:RTSS:09} shows how on Amazon EC2, a malicious client can obtain a virtual machine that shares the same physical server as the victim client with 40\% chance by just spending a few dollars; after this, the malicious client can launch side-channel attacks such as access CPU's data caches to obtain sensitive data from the client. 

We would like to mention that the cryptographic community has developed partially homomorphic encryption (PHE) systems in the past, where some operations can be performed on encrypted data. For example, any number of additions can be performed on data encrypted using Paillier cryptosystem~\cite{EUROCRYPT:Paillier:99}; any number of multiplications can be performed on data encrypted using RSA cryptosystem. Such PHE systems have been used as part of projects such as CryptDB~\cite{SOSP:PRZB:11} in combination with other approaches; however, PHE systems by themselves are not really useful as almost all real world queries would involve several operations that is beyond what can be handled by PHE systems.

Some approaches discuss having a trusted subcomponent on the service provider, which can process unencrypted data. A recent approach that uses a trusted subcomponent on the service provider is ~\cite{CIDR:ABE+:13}, where only the cardinality of the input and the output is leaked to the untrusted service provider; the untrusted service provider cannot obtain the size of intermediate results and other information. However, using a trusted component on the service provider appears to contradict with the multi-tenancy requirements of the cloud service provider.

We would therefore like the server be able to perform processing on encrypted data, and this has been the focus of research within the database community at least since 2002~\cite{SIGMOD:HILM:02}. While there have been significant advances for this problem~\cite{SIGMOD:HILM:02, SIGMOD:AKSX:04, EUROCRYPT:BCLO:09, SOSP:PRZB:11, CIDR:CJP+:11}, these approaches trade-off security guarantees with the processing capabilities of the service provider (see Section~\ref{sec:security}) for a detailed comparison).

In this paper, we investigate recent advances in fully homomorphic encryption (FHE)~\cite{STOC:Gentry:09} that enable computation of arbitrary functions on encrypted data, and how they can be used to provide the strongest security guarantees and the service provider can process any query. While FHE by itself enables any computation on encrypted data, using FHE naively for database systems is not practical. One of the difficulties is that FHE does not yield itself naturally to algebraic processing. Database systems have relied on algebraic processing for decades, where a query is compiled into a algebraic plan composed of different algebraic operations. In this paper, we provide solutions that enable algebraic processing of data encrypted using FHE, and outline several issues that need to be investigated before FHE based query processing systems can become practical. Our contributions in this paper are as follows:

\begin{itemize}

\item We compare various solutions proposed in literature with respect to two parameters: (a) security guarantees, and (b) query processing capabilities of the service provider. This comparison gives us a good idea of how the database community has progressed in this research area.

\item We provide a solution for algebraic processing of data encrypted using FHE. Our solution involves four main aspects: (a) a data model that is used for representing the relational tables (both the original tables and the intermediate results during algebraic query processing), (b) a computational model that can be used for query processing by the service provider on data encrypted using FHE, (c) algorithms for different relational algebra operators based on our data and computational models, and (d) techniques for transferring results of queries back to the client.

\item We outline several issues that still need to be studied before we can investigate the practicality of FHE based solutions for query processing.

\end{itemize}






\noindent {\bf Outline:} The outline for the rest of the paper is as follows. In Section~\ref{sec:problem}, we describe the problem and compare existing approaches in terms of how much of the query is processed by the service provider and the security guarantees. In Section~\ref{sec:models}, we describe the data model used in our approach, and the computational model that guarantees security while operating on encrypted data. In Section~\ref{sec:algebra}, we describe algorithms for various relational algbera operators and other essential operators in terms of our computational model; the input and output to the various relational algebra operators are tables represented in our data model. In Section~\ref{sec:sendResults}, we describe approaches for sending results to the client; Section~\ref{sec:issues} describes some of the issues that need more careful investigation into incorporating FHE based schemes for query processing; Section~\ref{sec:conclusions} concludes the paper.

\section{Problem Definition} \label{sec:problem}

In this paper, we consider two main aspects for enabling secure database as a service solutions: (a) what should be the capabilities of the service provider, and (b) what security guarantees should be provided.

\subsection{Role of the database service}
If a database service provider does not perform any query processing, then the data can be kept secure by encrypting the data on the client and using the service provider for only storing this encrypted data. In this case, any query processing will involve bringing the entire data from the service provider to the client and executing the query on the client. This defeats the purpose of using a database service provider. The cryptographic community prohibits such trivial solutions by requiring {\em compactness}~\cite{FOCS:Vaikuntanathan:11, EUROCRYPT:vGHV:10}, where the data sent back to the client as the result of query processing is required to be comparable to the unencrypted result size.  Note that the database service provider might still be able to perform operations such as project and count or even sum if data is encrypted using partially homomorphic encryption schemes and guarantee {\em compactness}, as these do not need access to unencrypted data; however, arbitrary selections, joins and almost all other operations cannot be performed by the service provider.



But then the question is: can the client participate in query processing? In most of the current solutions, the database service provider can perform only limited operations and the client performs the rest of the processing. In~\cite{SIGMOD:HILM:02}, the service provider can perform selection with false positives which need to be filtered at the client; OPE schemes~\cite{SIGMOD:AKSX:04, EUROCRYPT:BCLO:09} can perform range searches, but OPE schemes do not support other computations, and any left over query processing must be done at the client. In~\cite{DNIS:AEW:11}, a secure B+ tree index is dispersed across multiple servers in the cloud and the client performs the index traversal bringing in appropriate index nodes as needed from the cloud. 
A work that considers that the database service provider perform almost all the query processing is CryptDB~\cite{CIDR:CJP+:11, SOSP:PRZB:11}. In~\cite{CIDR:CJP+:11}, the client JDBC driver is responsible for encryption and decryption, as well as translating client queries to queries to be executed on the service provider on encrypted data. We will discuss CryptDB in more detail in Section~\ref{sec:security}. 

In this paper, we require that the service provider perform all the query processing, and the client is involved with encryption of data, encryption of literals in the query, and decryption of query results.

\subsection{What does ``secure'' mean?} \label{sec:security}

The cryptographic community~\cite{BOOK:KL:07} has defined several notions of security. We now examine these various notions of security 
and their implications to database as a service. 

Modern cryptography uses the notion of {\em semantic security}, which says that an encryption scheme is secure if no adversary that runs in polynomial time can learn any partial information about the plaintext from the ciphertext. An equivalent notion (that is used in practice) is {\em indistinguishability}: given a ciphertext $c$ that is an encryption of either $m_0$ or $m_1$, no adversary that runs in polynomial time can distinguish whether $c$ is an encryption of $m_0$ or $m_1$ better than a random guess, even when the adversary chooses $m_0$ and $m_1$. 
Also, most modern cryptographic constructions rely on {\em cryptographic assumptions} (typically about the hardness of well studied problems)~\cite{BOOK:KL:07}. For example, RSA public-key encryption is secure assuming that factoring is hard.

We now define who is the adversary in our model. A fairly conservative definition is that the database service provider is the adversary.
This requires that plaintext is available only on the client, and no partial information on the data is leaked if the database service provider is compromised~\cite{CCS:RTSS:09}, or to a curious DBA~\cite{SOSP:PRZB:11}. Furthermore, we often assume that our goal is confidentiality, and not integrity or availability~\cite{SOSP:PRZB:11}. In other words, the adversary will not modify queries, query results or stored data. For instance, if the client asks a query: get rows in table $R$ where $a > 5$, the adversary can modify the query to say: get rows in table $R$ where $a > 5 + 1$, the adversary can give partial results or delete some rows from a table; but will not perform these kinds of attacks. Other definitions have also been used; for instance,~\cite{VLDB:CKHM:10} assumes that a malware on the server is the adversary, but there is an additional trusted hardware security module (secure co-processor); here, the goal is to limit the plaintext data exposed to the malware.

We need to define the capabilities of the adversary; different capabilities of the adversary give different security definitions~\cite{BOOK:KL:07}. In chosen plaintext attack (CPA), the adversary can perform any number of encryptions. This definition is appropriate for public-key encryption systems, as the adversary knows the public key. In chosen ciphertext attack (CCA), the adversary is more powerful and can decrypt any ciphertext (other than the ``challenge ciphertext'').  If a database service provider adversary can decrypt ciphertexts, it may be difficult to ensure that the service provider is not decrypting sensitive data. We therefore believe that CPA security (and not CCA security) is appropriate for secure database service providers. Furthermore,~\cite{STOC:Gentry:09, FOCS:Vaikuntanathan:11} argue that FHE schemes that provide CCA2-security cannot exist.


CPA security generally requires that the encryption scheme is not deterministic. Deterministic encryption (DET) schemes cannot be used for 
small domains, and do not provide indistinguishability if values can repeat~\cite{BOOK:KL:07}. However, DET can provide CPA security if values do not repeat, for instance for keys in a (key, value) store. However, DET schemes appear not secure when query results are computed on encrypted data; for instance, if the count of values in a column yields a value in another column, is it secure? Order preserving encryption (OPE)~\cite{SIGMOD:AKSX:04, EUROCRYPT:BCLO:09} is weaker than DET and guarantees that order among plaintext values is preserved after encryption. This allows indexes to be built on ciphertexts that can be used for range searches. (Note that DET allows indexes to be built on ciphertexts that can be used for equality searches).

\begin{figure}[h]
\centering
\epsfig{file=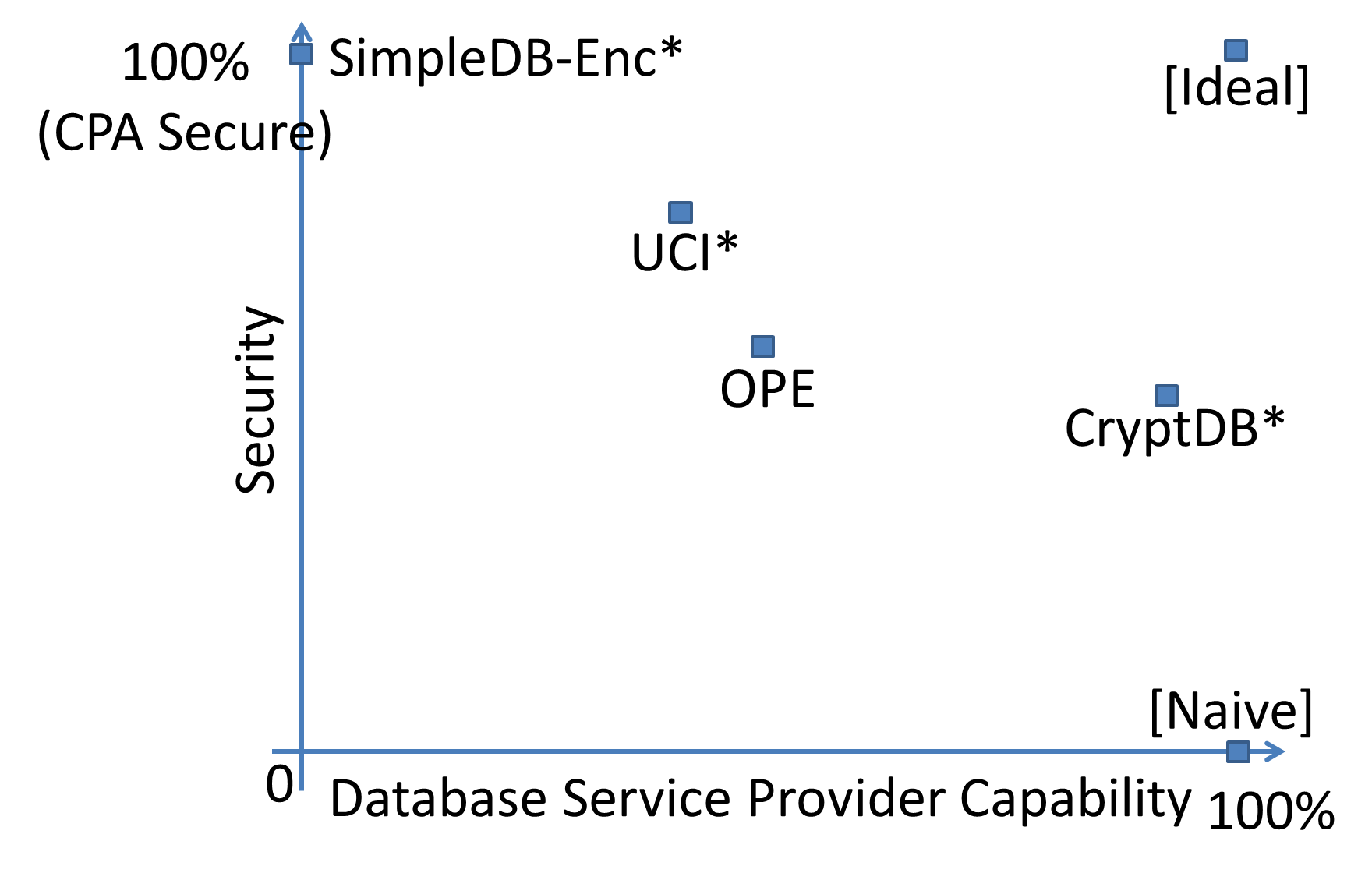, width=2.5in} 
\caption{Comparison of state of the art approaches} \label{fig:comparison}
\end{figure}

Figure~\ref{fig:comparison} compares some of the current approaches in terms of security and database service provider capabilities (we omit performance even though it is an important measure). We show SimpleDB where all data is encrypted; it is 100\% secure and provides no processing.
At the other extreme is [Naive] that stores plaintext; it has no security but can perform all query processing.
Our goal is [Ideal], with 100\% security and 100\% query processing. OPE approaches allow range searches, but are not CPA secure (note that encryption used by an OPE scheme will be deterministic). UCI (the approach in~\cite{SIGMOD:HILM:02}) provides a tunable bucketization. We get 100\% security with no query processing if all values in a column map to one bucket. In Figure~\ref{fig:comparison}, UCI* considers a bucketization where a few distinct values fall into one bucket: it has less capability than OPE, as false positives in the result from the service provider need to be filtered at the client.

The main idea behind CryptDB~\cite{SOSP:PRZB:11} is {\em adjustable security}, where each column is encrypted using multiple schemes into (multiple) onions. A column is brought to an appropriate encryption level depending on the operations to be performed: partially homomorphic encryption (such as Paillier) to compute sum, OPE for non-equality comparisons, RND encryption (which is CPA secure) if no operation is to be performed, etc. However, CryptDB cannot perform 100\% of the query processing on the service provider for two reasons: (a) minimum encryption for a column can be specified: for instance, if RND is minimum encryption for a column, then non-equality comparisons cannot be performed, (b) some composition of operations are prohibited, for instance, addition followed by a non-equality comparison is not secure. Also, as OPE-JOIN (range join) appears to have a weaker encryption than OPE, we show CryptDB with a weaker security than pure OPE schemes. 

We can now define the problem that we would like to solve in order to enable secure database as a service solutions: provide solutions that provide CPA-security (denoted 100\% CPA-secure solution in Figure ~\ref{fig:comparison}), and where the service provider performs all the query processing. In other words, the client is involved only with encryption of data and query literals, and with decryption of query results.



\section{Data and Computational Models} \label{sec:models}

In this section, we introduce our data and computational models. The data model describes how the input to any query operator is represented. As operators can be composed, the result of any query operator is also represented using the same data model. The computational model describes what computations are permitted: remember that while the data is encrypted and the query literals are encrypted, some unencrypted data may also be needed during query processing. The data and the computational models we present will enable algebraic query processing on encrypted data. However, before we describe these models, we would like to introduce fully homomorphic encryption (FHE) as defined by the cryptographic community.

\subsection{Fully Homomorphic Encryption (FHE)}

A homomorphic encryption scheme consists of an $\eval$ algorithm in addition to the KeyGen, Encrypt and Decrypt algorithms that are part of any encryption scheme. The $\eval$ algorithm can evaluate any of a set of ``permitted'' circuits, and produces ``correct'' and {\em compact} ciphertexts as output. A fully homomorphic encryption (FHE) scheme is homomorphic for all circuits. The main technique in~\cite{STOC:Gentry:09} is {\em bootstrapping}, where a somewhat homomorphic encryption scheme (that is, one for which $\eval$ is possible only for a few small circuits), is used to make a fully homomorphic encryption scheme. 
Note that in order to encrypt, a small amount of noise is added, and the more complicated the circuit, the more this noise is amplified when $\eval$ is applied to the ciphertext.  When the noise grows too large, the output of $\eval$ is no longer correct.  Bootstrapping periodically uses $\eval$ to evaluate the decryption circuit itself.  This re-encrypts the ciphertext, and ``refreshes" the amount of noise. The above scheme gives us {\em leveled FHE} that can evaluate any circuit of depth at most $d$ (where we refresh at most $d$ times). If we further assume ``circular security" (that is, it is safe to make public the FHE secret key encrypted using its own public key), then we get a ``pure" FHE that can truly evaluate all circuits~\cite{FOCS:Vaikuntanathan:11}. In~\cite{STOC:Gentry:09, FOCS:Vaikuntanathan:11}, the permitted circuits in Evaluate include the universal gates \{XOR, AND\}.

The schemes based on Craig Gentry's original scheme are very time consuming. For evaluating one ``gate'', (a circuit that requires a single refresh) the running time is $\Omega(\lambda^4)$, where $\lambda$ is the security parameter. Since~\cite{STOC:Gentry:09}, various constructions for FHE have been proposed which 
make better cryptographic assumptions and the per-gate running time has been decreased to $O(\lambda)$~\cite{FOCS:Vaikuntanathan:11}. It is worth noting that even though such advances are promising, practicality of FHE schemes is still unclear.

Conceptually, an FHE scheme can evaluate any function, and hence can process any query. However, applying FHE to query processing is not straightforward, and several issues have to be addressed. The first one is how to translate any query into a FHE circuit that can be evaluated on encrypted data. The database community has long used relational algebraic processing to execute queries, where any query can be translated into a plan consisting of a set of relational algebra operators. We study how to perform algebraic processing on data encrypted using FHE schemes.

\begin{figure}[h]
\centering
\epsfig{file=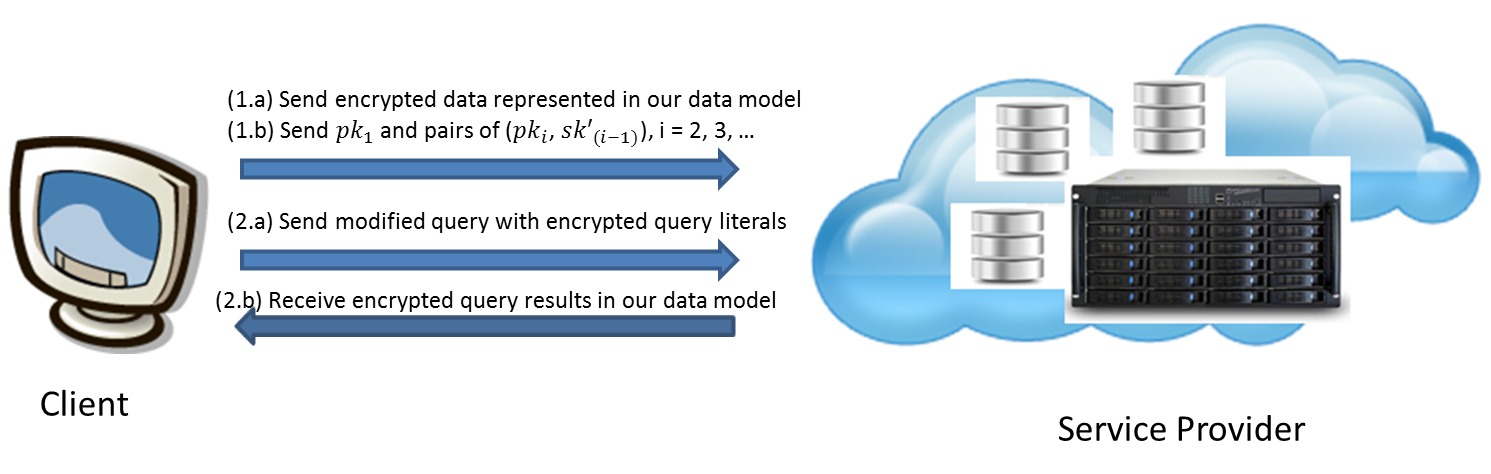, width=5in} 
\caption{Proposed Architecture for Processing Encrypted Data} \label{fig:arch}
\end{figure}

Our architecture for processing encrypted data is shown in Figure~\ref{fig:arch}. Here step 1.a is part of the initial setup, where the client sends encrypted data to the service provider. Step 1.b is also part of the initial setup, here a sequence of keys as needed for bootstrapping are sent to the service provider as follows: the client sends $(pk_2, sk_1')$; $pk_2$ is the second public key, $sk_1'$ is the first secret key ($sk_1$) encrypted using the second public key $pk_2$; similarly the client sends $(pk_3, sk_2')$ and so on, depending on the depth of the circuits that can be executed by $\eval$. Recall that if we assume circular security, the client needs to send the service provider only $(pk, sk')$, where $sk'$ is $sk$ encrypted using $pk$. Steps 2.a and 2.b show the query processing; In Step 2.a, the query is modified to encrypt any literals in the query; this modified query is sent to the service provider that processes the query (using the operators as discussed in Section~\ref{sec:algebra}); the encrypted query results will be sent back to the client (as described in Section~\ref{sec:sendResults}. The client can decrypt the query results using its available secret keys.

\subsection{Data Model}

For relational algebraic processing, the operands are tables, and the result of any operator is a table. This result table can in turn be used as an operand by another operator. However, this does not work naively for encrypted data. We propose a simple extension to the data model as follows. 

Suppose the unencrypted table is $R(A)$, where $R$ is the name of the table and $A$ is the set of columns in $R$. This table is represented in our model as $(R', pk)$, where $pk$ is the public key used to encrypt the values in the table (the granularity of encryption is a value or a cell; also all the values in a table are encrypted using the same key). The table $R'(A, p)$ has all the columns in $A$ and an additional presence column $p$ that indicates the presence of the row in the table; $p$ takes values $0$ or $1$ (the $p$ value will be encrypted using $pk$ as well). This model is used for representing any base table as well as any intermediate table during query processing. In Section~\ref{sec:algebra}, we define algorithms for relational algebra operators that take operands in this model and produce results in this model as well.

An example table is shown in Table~\ref{tab:example1}. Here $\overline{x}$ represents $x$ encrypted using the public key (say $pk$). The presence bits indicate the first two rows are present ($p = \overline{1}$), and the third row is not present in the table ($p = \overline{0}$). Note that the structure of the database is known to the server, only the individual values are encrypted. Also the server knows an upper bound of the number of rows in each table (for the table below, this upper bound is 3).

\begin{table}[ht]
\centering
\begin{tabular}{l|l|l|l|l|l}
model & speed & ram & hd & price & p \\
\hline
\hline 
$\overline{1001}$ & $\overline{3}$ & $\overline{1024}$ & $\overline{250}$ & $\overline{2114}$ & $\overline1$ \\
\hline
$\overline{1002}$ & $\overline{2}$ & $\overline{512}$ & $\overline{80}$ & $\overline{478}$ & $\overline1$ \\
\hline
$\overline{1003}$ & $\overline{1}$ & $\overline{512}$ & $\overline{250}$ & $\overline{600}$ & $\overline0$ \\
\hline
\end{tabular}
\caption{Example Table in our Data Model}
\label{tab:example1}
\end{table}

\subsection{Computational Model}

We present a computational model below that can operate on encrypted and unencrypted data, limiting the constructs that can be applied to encrypted data. In Section~\ref{sec:algebra}, we will develop algorithms for the various relational algebra operators using this computational model. Our computational model consists of the following:

\begin{itemize}
\item unencrypted literals can be assigned to variables, and all programming language control structures can be used on these variables and literals
\item encrypted literals can be assigned to variables, but no programming language control structure can be used on these variables and literals
\item service provider can make function calls with encrypted or unencrypted literals and variables as parameters
\item service provider can obtain encryption of any literal using a specified key (the encryption keys are public)
\item service provider can evaluate any fixed, combinatorial circuit on encrypted values using $\eval$
\end{itemize}

Let us consider a couple of examples to illustrate how to manipulate data using the above computational model. The first example is to assign the larger of values in two variables, $a$ and $b$ to the variable $x$. If $a$ and $b$ are unencrypted, one algorithm using the above computational model is shown on the left. However, suppose $a$ and $b$ are encrypted. The algorithm on the left is not allowed by our computational model, as we cannot write an if construct using a comparison on encrypted data. However, the algorithm on the right can now be used. Here, see that $a > b$ can be performed using a fixed, combinatorial circuit using $eval$; the resulting encrypted value can be assigned to a variable, $flag$. Similarly, the expression $(flag * a) + (NOT(flag) * b)$ can be evaluated using a fixed, combinatorial circuit, and the resulting encrypted value can be assigned to variable $x$.
\\

\begin{center}
\begin{tabular}{ll}
\begin{minipage}[c]{2in}
\begin{verbatim}
if (a > b) 
  x = a;
else x = b;
\end{verbatim}
\end{minipage}
&
\begin{minipage}[c]{2in}
\begin{verbatim}
flag = a > b;
x = (flag * a) +
       (NOT(flag) * b);
\end{verbatim}
\end{minipage}
\end{tabular} \\
\end{center}

A second example is that the server can perform loops with fixed ranges using an unencrypted loop counter, as long as the terminating condition is not dependent on any encrypted data. For instance, the server can iterate over the ``3'' rows in Table~\ref{tab:example1} (as the server knows the number of rows in the table). 

Any database operation that can be described as compositions of the above components can be implemented securely with FHE. The security of this computational model follows directly from the security of the FHE scheme, as manipulation of encrypted values are performed by the service provider only using $\eval$.

\section{Operator Algorithms} \label{sec:algebra}

We will first define basic operators, and then we will build other operators using these basic operators.

\subsection{Bitwise Operators}

In~\cite{EUROCRYPT:vGHV:10}, the authors describe how AND, XOR and refresh can be evaluated. Remember that AND, XOR form a set of universal gates in Boolean logic; hence any computer program can be expressed using these two gates.  

\noindent
$XOR$: denoted $(b_1, pk_1) \bigoplus (b_2, pk_1) \rightarrow (b, pk)$. \\
$AND$: denoted $(b_1, pk_1) \bigotimes (b_2, pk_1) \rightarrow (b, pk)$. \\

\noindent
Here $(b_1, pk_1)$ indicates that $b_1$ is the encrypted (using public key $pk_1$) value of a bit (similarly $(b_2, pk_1)$). The result of XOR/AND is $(b, pk)$, which is encrypted using $pk$; $pk$ could be $pk_2$ (the next public key) if the errors in $b_1$ and/or $b_2$ necessitate bootstrapping; if no bootstrapping is required, $pk = pk_1$.  \\

\noindent We extend the bitwise operators to allow operands encrypted using different keys. 

\noindent
$XOR$: $(b_1, pk_i) \bigoplus (b_2, pk_{i+j}) \rightarrow (b, pk)$.

\noindent
This is performed by first bootstrapping $b_1$ to obtain $b_1'$, which is $b_1$ encrypted using $pk_{i+j}$. We now perform $(b_1', pk_{i+j}) \bigoplus (b_2, pk_{i+j})$ to obtain $(b, pk)$. Similarly we define $AND$ as:

\noindent
$AND$: $(b_1, pk_i) \bigotimes (b_2, pk_{i+j}) \rightarrow (b, pk)$. \\

\noindent We can define other Boolean operators using $XOR$ and $AND$ as in any logic design textbook~\cite{BOOK:MK:08}.

\noindent $NOT$: $NOT(b_1, pk_i) \rightarrow (b, pk)$ $=$ $(b_1, pk_i) \bigoplus (\overline{1}, pk_i)$

\noindent $OR$: $(b_1, pk_i) \ OR \ (b_2, pk_{i+j}) \rightarrow (b, pk)$ $=$  $((b_1, pk_i) \bigotimes (b_2, pk_{i+j}))$ $\bigoplus$ $((NOT(b_1, pk_i) \bigotimes (b_2, pk_{i+j}))$ $\bigoplus$ \\
\hspace*{2.5in} $((b_1, pk_i) \bigotimes NOT(b_2, pk_{i+j})))$ \\

\noindent We often drop the key information when we represent the operations, and just say $x_1$ $OR$ $x_2$.

\subsection{Arithmetic and Comparison Operators}
\label{sec:ArithmeticComparisonOps}

In this section, we will examine addition (other arithmetic operators can be defined similarly) and basic comparison operators (equality and greater than). Other comparison operators can be defined using these two operators and using and, or and not. \\

\noindent Addition: $(x_1, pk_1) \boxplus (x_2, pk_1) \rightarrow (x, pk)$. \\
The addition operator takes two numbers $x_1$ and $x_2$ encrypted using the same key, performs a full adder as described in~\cite{BOOK:MK:08}. 
The result is encrypted using $pk$; $pk$ can be $pk_1$ or a later public key if bootstrapping is performed (same is true for other operators in this section as well). \\

\noindent Equality: $(x_1, pk_1) \doteqdot (x_2, pk_1)$ $\rightarrow$ $(b, pk)$.~\cite{DBKDA:GGE:11} \\
The equality comparison operator takes two values $x_1$ and $x_2$ encrypted using the same key, and returns an encrypted bit; the result bit is $1$ if the values are equal, and $0$ if the values are not equal. 
The algorithm is given below. Let $x_1$ and $x_2$ represent $n$ bit values; let the encrypted values for the bits be $x_{11}, x_{12}, \ldots, x_{1n}$ and $x_{21}, x_{22}, \ldots, x_{2n}$ respectively. \\

\begin{tabular}{cc}
\noindent
\fbox{
\parbox{0.36\linewidth}{
\noindent
$result = (\overline{1}, pk_1)$ // encryption of 1 using $pk_1$ \\
for $i = 1 \ldots n$ \\
\mySpOne $temp = (x_{1i} \bigoplus x_{2i}) \bigoplus \overline{1}$ \\
\mySpOne // $temp = 1$ if $x_{1i} = x_{2i}$; else $0$ \\
\mySpOne $result = result \bigotimes temp$ \\
return $b = result$. 
}}
&
\noindent
\fbox{
\parbox{0.5\linewidth}{
\noindent
$result = (\overline{0}, pk_1)$; $done = (\overline{0}, pk_1)$; \\
for $i = 1 \ldots n$ \\
\mySpOne $t_1 = x_{1i} \bigotimes NOT(x_{2i})$; $t_2 = x_{2i} \bigotimes NOT(x_{1i})$; \\
\mySpOne $result = (done \bigotimes result) \bigoplus (NOT(done) \bigotimes t_1)$; \\
\mySpOne // keep result if already done; otherwise set to $t_1$ \\
\mySpOne $done = done \bigoplus (NOT(done) \bigotimes (t_1 \ OR \ t_2))$; \\
return $b = result$. 
}}
\\
Equality & Greater Than
\end{tabular} \\


\noindent Greater Than: $(x_1, pk_1) \gtrdot (x_2, pk_1)$ $\rightarrow$ $(b, pk)$. \\
The greater than comparison operator takes two values $x_1$ and $x_2$ encrypted using the same key, and returns an encrypted bit; the result bit is $1$ if $x_1 > x_2$, and $0$ otherwise.
Just like for equality comparison, assume $n$ bits for $x_1$ and $x_2$.
\\


\noindent All the operators can be extended (using bootstrapping) to the case when the two operands are encrypted using different keys.

\subsection{Operations on a Bit and a Word}

In some cases, we may want to operate on a bit and and a number. We will see later that as part of COUNT, we add a number and a bit; as part of SUM, we perform AND of a number and a bit. 

Suppose $b$ is an encrypted bit, and $x$ is a number with $n$ bits, say $x_1, x_2, \ldots, x_n$, then $x \bigotimes b$ $=$ $x_1', x_2', \ldots, x_n'$, where $x_i' = x_i \bigotimes b$. Similarly $x \boxplus b$ $=$ $x \boxplus b_{num}$, where $b_{num} = x_1', x_2', \ldots x_n'$, where $x_i' = \overline{0}$ (i.e., encryption of 0), for $1 \leq i <  n$, and $x_n'= b$.


\subsection{Implementing Relational Algebra} \label{sec:opAlgms}

We now describe our algorithms for the different relational algebra operators in terms of the operators and the computational model defined earlier. \\

\noindent {\bf Select} $\sigma_c (R', pk_i) \rightarrow (R'', pk)$

\noindent The select operator takes a selection condition $c$, an input table $R'$ encrypted using $pk_i$ and produces a result table $R''$ encrypted using $pk$. The schema for both $R'$ and $R''$ are $(A, p)$, where $p$ is the encrypted presence bit. Note that the selection condition $c$ can be represented as a combinatorial circuit using and, or, not and the comparison operators (Section~\ref{sec:ArithmeticComparisonOps}) (we do not consider {\tt LIKE} in this work); we consider $c$ as a function $c(r_x) \rightarrow b$ that takes a row of $R'$ and returns an encrypted bit with value $1$ if the row satisfies the condition in $c$; value $0$ if the row does not satisfy the condition. 
The algorithm for select is given below:

\begin{tabular}{cc}
\noindent
\fbox{
\parbox{0.4\linewidth}{
\noindent
for each row $r_x = (a_x, p_x)$ in $R'$ \\
\mySpOne add row $(a_x, p_x')$ to $R''$; $p_x' = p_x \bigotimes c(a_x)$ \\
bootstrap all values in $R''$ as required
}} &
\noindent
\fbox{
\parbox{0.4\linewidth}{
\noindent
for each row $r_x = (a_x, p_x)$ in $R'$ \\
\mySpOne add row $\pi_L(a_x), p_x$ to $R''$}}
\\
Select & Project
\end{tabular} \\

\noindent {\bf Project} $\pi_L (R', pk_i) \rightarrow (R'', pk)$

\noindent The project operator takes an input table $R'(A, p)$, a set of attributes $L (L \subseteq A)$ and produces a result table $R''(L, p)$ (i.e., keeps only the columns $L$).  \\

\noindent {\bf Cross Product}: $(R_1', pk_i) \times (r_2', pk_j)$ $\rightarrow$ $(R'',pk)$

\noindent This operator takes two input tables $R_1'(A_1, p)$ and $R_2'(A_2, p)$; the output table schema is $R''(A_1, A_2, p)$. \\

\begin{tabular}{cc}
\noindent
\fbox{
\parbox{0.4\linewidth}{
\noindent
for each row $r_{1x} = (a_{1x}, p_{1x})$ in $R_1'$ \\
\mySpOne for each row $r_{2x} = (a_{2x}, p_{2x})$ in $R_2'$ \\
\mySpTwo add row $(a_{1x}, a_{2x}, p_{1x} \bigotimes p_{2x})$ to $R''$ \\
bootstrap all values in $R''$ as required.
}} &
\noindent
\fbox{
\parbox{0.35\linewidth}{
$count = (\overline{0}, pk_i$)  // encryption of 0 \\ 
for each row $r_{x} = (a_{x}, p_{x})$ in $R'$ \\
\mySpOne $count = count \boxplus p_x$; \\
return $count$}} \\
Cross Product & Count
\end{tabular} \\

\noindent {\bf Count} $COUNT_c(R', pk_i)$ 
$\rightarrow$ $(v, pk)$

\noindent The count function takes as input $R'(A, p)$ and returns the number of rows in the table $R'$ with 1 for the presence bit (null values are ignored in this work, $COUNT_c(R', pk_i)$ $=$ $COUNT_*(R', pk_i)$). \\

\noindent {\bf Sum} $SUM_c(R', pk_i) \rightarrow (v, pk)$ (similar to count).


\begin{tabular}{cc}
\noindent
\fbox{
\parbox{0.32\linewidth}{
\noindent
$sum = (\overline{0}, pk_i)$  // encryption of 0 \\ 
for each row $r_{x} = (a_{x}, p_{x})$ in $R'$ \\
\mySpOne $sum = sum \boxplus (\pi_c(r_x) \bigotimes p_x)$; \\
return $sum$ 
}} &
\noindent
\fbox{
\parbox{0.45\linewidth}{
\noindent $found = (\overline{0}, pk_i)$; // is min valid \\
for each row $r_{x} = (a_{x}, p_{x})$ in $R'$ \\
\mySpOne $f = p_x \bigotimes ((found \bigotimes (\pi_c(a_x) \gtrdot min))$ $\bigoplus$ \\
\mySpTwo   $NOT$$(found))$ // do we have a new min? \\
\mySpOne $found = found \bigoplus (NOT(found) \bigotimes p_x)$ \\
\mySpOne $min = (f \bigotimes \pi_c(a_x)) \bigoplus (NOT(f) \bigotimes min)$ \\
return $min$}} \\
Sum & Min
\end{tabular} \\

\noindent {\bf Min} $MIN_c(R', pk_i) \rightarrow (v, pk)$

\noindent Max is similar to Min; Average can be computed using Sum and Count using circuits for division~\cite{BOOK:MK:08}. \\

\noindent {\bf Distinct} $\delta(R', pk_i) \rightarrow (R'', pk)$ 

\noindent The distinct operator takes an input table $R'(A, p)$ and produces a result table $R''(A, p)$ without any of the duplicate rows in $R'$
(i.e., $p$ is set to $\overline{0}$ for duplicate rows). \\

\begin{tabular}{cc}
\noindent
\fbox{
\parbox{0.4\linewidth}{
\noindent
Copy $R'$ to $R''$ \\
Let $n$ be the number of rows in $R'$ (also $R'')$ \\
for $i = 2 \ldots n$ \\
\mySpOne $f= \overline{0}$ // is row $i$ a duplicate? \\
\mySpOne for $j = 1 \ldots (i - 1)$ \\
\mySpTwo $equals = (a_i \doteqdot a_j) \bigotimes p_j$; \\
\mySpTwo $f = f \bigoplus (NOT(f) \bigotimes equals)$ \\
\mySpOne set $p_i$ in $R''$ = $(f \bigotimes \overline{0}) \bigoplus (NOT(f) \bigotimes p_i)$ \\
bootstrap all values in $R''$ as required.
}} &
\begin{tabular}{l}
\noindent
\fbox{
\parbox{0.4\linewidth}{
\noindent
Copy $R'$ to $R''$. \\
for $i = 1 \ldots n$ \\
\mySpOne for $j = 1 \ldots (n - 1)$ \\
\mySpTwo $f = \pi_c(r_j) \gtrdot \pi_c(r_{j+1})$ \\
\mySpTwo swap $(r_j, r_{j+1}, f)$ }} \\
\fbox{
\parbox{0.4\linewidth}{
swap $(x, y, f)$ \\
\mySpOne $t_1 = x$; $t_2 = y$; \\
\mySpOne $x = (f \bigotimes t_2) \bigoplus (NOT(f) \bigotimes t_1)$ \\
\mySpOne $y = (f \bigotimes t_1) \bigoplus (NOT(f) \bigotimes t_2)$
}}
\end{tabular}
\\
Distinct & Sort
\end{tabular} \\

\noindent {\bf Sort} $\tau_L(R', pk_i) \rightarrow (R'', pk)$ 

\noindent
The sort operator sorts the rows in $R'$ based on $L$.
For simplicity, we will illustrate sorting on 1 column $c$ in ascending order; it can be easily extended. \\

\noindent {\bf Group By}: $\gamma_{L, AGG}(R', pk_i) \rightarrow (R'', pk)$. 

\noindent The group by operator groups the rows in $R'$ based on values in $L$ columns, and computes the aggregate functions in $AGG$ for each group. For simplicity, we will consider only one SUM aggregate function; multiple aggregates can be computed similarly.

\begin{center}
\noindent
\fbox{
\parbox{0.8\linewidth}{
\noindent
$\tau_L(R', pk_i) \rightarrow (R''', pk_j)$. Let there be $n$ rows in $R'''$: $(a_1, p_1)$, $(a_2, p_2)$, $\ldots$, $(a_n, p_n)$. \\
$sum = \overline{0}$ // starting sum \\
$f = \overline{0}$ // current group has $> 0$ elements? \\
for $i = 1 \ldots n$ \\
\mySpOne if (i != 1) \\
\mySpTwo $f_1 = (\pi_L(pr) \doteqdot \pi_L(r_i))$; // $f_1 = 1$ means add to prev sum \\
\mySpTwo Add $(\pi_L(pr), sum, NOT(f_1) \bigotimes f)$ to $R''$. \\
\mySpTwo $f = (NOT(f_1) \bigotimes p_i) \bigoplus (f_1 \bigotimes (f \ OR \ p_i))$ \\
\mySpOne else $f_1 = \overline{0}$; $f = p_i$; \\
\mySpOne $pr = r_i$; \\
\mySpOne $v = (p_i \bigotimes \pi_c(r_i)) \bigoplus (NOT(p_i) \bigotimes \overline{0})$; // $v$ is set to current $c$ value or $0$ \\
\mySpOne $sum = ((f_1 \bigotimes sum) \bigoplus (NOT(f_1) \bigotimes \overline{0})) \boxplus v$ \\
Add a row $\pi_L(pr), sum, f$ to $R''$. // last group 
}} \\
\end{center}

\noindent {\bf Bag Union} $(R_1', pk_i) \sqcup (R_2', pk_j)$ $\rightarrow$ $(R'', pk)$ (the schema for the 2 input tables and the result table is $(A, p)$).

\begin{center}
\fbox{
\parbox{0.45\linewidth}{
\noindent
add each row $r_{1x}$ in $R_1'$; 
add each row $r_{2x}$ in $R_2'$ \\
bootstrap all values in $R''$ as required.
}}
\end{center}

\noindent {\bf Bag Intersection} $(R_1', pk_i) \sqcap (R_2', pk_j)$ $\rightarrow$ $(R'', pk)$ (the schema for all the 3 tables is $(A, p)$). \\

\begin{center}
\noindent
\fbox{
\parbox{0.86\linewidth}{
\noindent
$\tau_A(R_1') \rightarrow R''$; $\tau_A(R_2') \rightarrow R_2''$; // $R''$ has sorted $R_1$ rows. $R_2''$ has sorted $R_2'$ rows. \\
// $R'' = (a_{11}, p_{11})$, $(a_{12}, p_{12})$, $\ldots$, $(a_{1{n_1}}, p_{1{n_1}})$. $R_2''$ $=$ $(a_{21}, p_{21})$, $(a_{22}, p_{22})$, $\ldots$, $(a_{2{n_2}}, p_{2{n_2}})$ \\
$i_2 = e^1$; // index for $R_2''$, starting with row 1 \\
for $i = 1 .. n_1$ \\
\mySpOne $f = \overline{0}$; // is current row in $R''$ matched? \\
\mySpOne for $j = 1 .. n_2$ \\
\mySpTwo $f_1 = i_2 \gtrdot j$; // used to skip rows in $R_2''$ \\
\mySpTwo $eq = a_{1i} \doteqdot a_{2j}$; $gt = a_{1i} \gtrdot a_{2j}$; \\
\mySpTwo $f_2 = (NOT(f) \bigotimes NOT(f_1) \bigotimes eq \bigotimes p_{1i} \bigotimes p_{2j})$; // result p bit must be 1 \\
\mySpTwo $f = (f \ OR \ f_2)$; \\
\mySpTwo $i_2 = i_2 \boxplus (f_2 \ OR \ gt \ OR \ NOT(p_{2j}))$; \\
\mySpOne $p_{1i} = f$; \\
bootstrap all values in $R''$ as required.
}} \\
\end{center}

\noindent The algorithm for {\bf Bag Difference} is same as that for Bag Intersect, except that for each outer loop, we set $p_{1i} = NOT(f) \bigotimes p_{1i}$.

\section{Returning Results to the Client} \label{sec:sendResults}

So far, we have seen how the service provider can compute the results of a query in a {\em provably secure} manner. Now the results need to be sent back to the client. Sending the entire table that is the output of the last relational algebra operator in the plan will violate the compactness requirement (Section~\ref{sec:problem} as this table is likely to be very large in size compared to the actual result size. 
Gentry~\cite{CACM:Gentry:10} remarks that the client needs to specify the size of the output, as the service provider cannot determine this size without knowing some relationship between the function to be evaluated and the data. 
In~\cite{UCSB:WAE:12}, the client specifies the number of result rows he/she wants as $n$, based on an estimation of the result size. The service provider packs $n$ rows; each is a valid result row with high probability. However, this might result in approximate query results as some results may not be sent to the client. 

\begin{figure}[h]
\centering
\epsfig{file=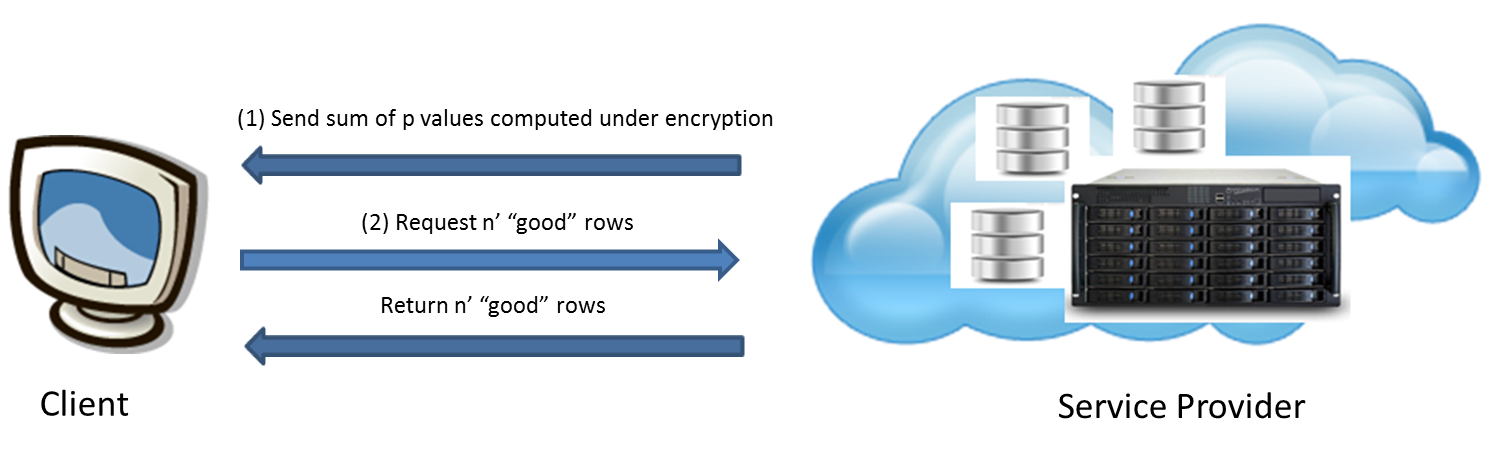, width=5in} 
\caption{Sending Query Results to the Client} \label{fig:sendResults}
\end{figure}

We propose a two-step process for sending the results to the client. First, the service provider computes the sum of the $p$ column values in the result table underneath the encryption. This is sent to the client; the client decrypts the sum to a value, say $n$. The client asks for $n'$ rows from the service provider (where $n' \geq n$, and the service provider can see $n'$). The service provider sorts the rows in the result table on the $p$ values underneath the encryption, while also maintaining any other ordering (specified in the query) and sends the ``top'' $n'$ rows. The client can verify that the sum of the $p$ column values equals $n$ (the service provider never knew $n$). This two-step process provides compact results and exact results back to the client. This process is shown in Figure~\ref{fig:sendResults}.

Note that the security of this two-step process needs to be studied more carefully. What if the service provider sent a sensitive value as the number of rows $n$, and when the client sends $n'$, the service provider knows an upper bound on the sensitive value. However, note that such an attack can be immediately detected by the client, by checking that the number of good rows in the result matches $n$. Another approach with provable security, but where the client receives approximate query results is discussed in~\cite{UCSB:WAE:12}.

\section{Research Issues} \label{sec:issues}

We list some of the key research issues that need to be investigated to make FHE based query processing a reality, and thus enable secure database as a service.

{\em Practicality of FHE} is a very important concern. However, recent advances are promising~\cite{FOCS:Vaikuntanathan:11}:
evaluating larger circuits without bootstrapping is described in~\cite{FOCS:Vaikuntanathan:11}; more efficient bootstrapping is discussed in~\cite{PKC:GHS:12}.

We need to understand the {\em implications of the computational model}, such as the time complexity for problems in query processing. 
Gentry~\cite{CACM:Gentry:10} 
remarks that random access speedups cannot work if the data is encrypted, and any algorithm must have a running time at least linear in the number of inputs. For instance, binary search on an ordered list of $t$ encrypted items will take much longer than $O(log \ t)$.
This is crucial for developing ``efficient'' operator implementations. Also {\em indexes} are essential for database system performance, and provide sub-linear running times. We need ingenious solutions that investigate indexes for FHE schemes.

We need to understand the properties of the operator algorithms and their impact on {\em query optimization}. For instance, current optimizations use heuristics such as selection push-down that is based on the property that the result of selection is likely to have fewer rows than the input table. However, the result of the select algorithm we describe in Section~\ref{sec:opAlgms} will have the same number of rows as in the input table. Therefore, selection push-down may not be useful.

We need to study the impact of FHE schemes on {\em cost-based optimization}. For this, we need to study what is a good cost model. 
In current systems, the costliest operation is (typically) accessing data on disk.
However, what is the costliest operation for query processing using FHE schemes: is it $\eval$ or bootstrapping or disk access? Studying such costs will allow us to build a good cost model that can be used for effective cost-based optimization.


A database server provides many more {\em functionalities} than processing SELECT statements, including processing update statements, user privileges and authentication, transactions, concurrency control, crash recovery etc. Further, the service provider needs to manage many types of {\em schema objects} such as views, indexes, stored procedures, triggers, user defined types etc. Some of these are discussed in~\cite{SOSP:PRZB:11}. There are several interesting questions: for instance, who handles user authentication when the service provider is untrusted? 


\section{Conclusions} \label{sec:conclusions}
In this paper, we described some of the approaches to enable secure database as a service, and how FHE schemes can contribute to this effort. 	FHE schemes are promising as they can provide the highest level of security, while the database service provider can evaluate the complete query, unlike current schemes in literature where the client participates actively in query processing. We also based our approach on algebraic processing, and provided algorithms for the various relational algebra operators based on the computational model that is proven to be secure.

There are several exciting research questions based on the new computational model, including practicality of FHE schemes,
building indexes, user authentication etc. We believe that using FHE to enable database as a service is ripe for significant involvement from the database community.

\vspace*{10pt}
\noindent
{\bf Acknowledgments} to several colleagues at University of Michigan, especially Mary Wootters and to the encouragement from several database researchers.

%
\bibliographystyle{abbrv}
\bibliography{cloud}  
%
%
\end{document}